\documentclass[useAMS,usenatbib]{mn2e} 
\usepackage{graphicx,lscape}
\def\integral{{\it INTEGRAL}} 
\def\rxte{{\it RXTE}} 

\def\chandra{{\it Chandra}} 
\def \gx {GX~339--4} 

\def \lerg {erg s$^{-1}$}

\title[State transitions of GX 339--4]{Broad-band X-ray spectral evolution of \gx\ during a state transition \thanks{Based on
observations with {\it INTEGRAL}, an ESA  project with instruments and science data centre funded by ESA member states 
(especially the PI countries: Denmark, France, Germany, Italy,  Switzerland, Spain), Czech
Republic and Poland, and with participation of Russia and the USA.}}
\author[M. Del Santo et al.]{M. Del Santo$^{1}$, T.M. Belloni$^{2}$, J. Homan$^3$, A. Bazzano$^{1}$, P. Casella$^4$, R.P. Fender$^5$, 
\newauthor E. Gallo$^6$, N. Gehrels$^7$, W.H.G. Lewin$^3$, M. M\'endez$^8$, M. van der Klis$^4$ \\ 
$^{1}$INAF/Istituto di Astrofisica Spaziale e Fisica Cosmica - Roma, Via Fosso del Cavaliere 100, I-00133 Roma, Italy\\
$^{2}$INAF/Osservatorio Astronomico di Brera, Via E. Bianchi 46, I-23807 Merate (LC), Italy\\ 
$^3$Center for Space Research, Massachusetts Institute of Technology, 77 Massachusetts Avenue, Cambridge, MA 02139-4307, USA\\ 
$^4$Astronomical Institute 'Anton Pannekoek', University of Amsterdam, and Center for High Energy Astrophysics, Kruislaan 403,\\ 
1098 SJ, Amsterdam, the Netherlands\\ 
$^5$School of Physics and Astronomy, University of Southampton, Southampton, Hampshire, SO17 1BJ\\ 
$^6$Physics Department, Broida Hall, University of California, Santa Barbara, CA 93106, USA\\ 
$^7$NASA Goddard Space Flight Center, Greenbelt, MD 20771, USA\\ 
$^8$Kapteyn Astronomical Institute, University of Groningen, P.O. Box 800, 9700 AV Groningen, the Netherlands}

\begin{document}

\date{Accepted ... Received ...}

\pagerange{\pageref{firstpage}--\pageref{lastpage}} \pubyear{0000}

\maketitle

\label{firstpage}

\begin{abstract}
We report on X-ray and soft $\gamma$-ray observations of the black-hole
candidate \gx\ during its 2007 outburst, performed with the \rxte\ and
\integral\ satellites. The hardness-intensity diagram of all \rxte/PCA data
combined shows a {\it q}-shaped track similar to that observed in previous
outbursts.The evolution in the diagram suggested that a  
transition from hard-intermediate state to soft-intermediate
state occurred, simultaneously with \integral\ observations performed in March.
The transition is confirmed by the timing analysis presented in this work, which
reveals that a weak type-A quasi-periodic oscillation (QPO) replaces a strong type-C QPO.
At the same time, spectral analysis shows that the flux of the high-energy component shows a significant decrease in its
flux. However, we observe a delay (roughly one day) between variations of the spectral parameters of the high-energy component
and changes in the flux and timing properties.
The changes in the high-energy component can be explained either in terms the high-energy cut-off
or in terms of a variations in the reflection component. 
We compare our results with those from a similar transition during the 2004
outburst of \gx.
\end{abstract}


\section{Introduction}

Since its discovery \cite{markert73}, the black-hole candidate (BHC)
\gx\ has been observed to spend long periods in outburst. Although
historically it was found to be mainly in the hard state (Maejima et al.
1984; Ilovaisky et al. 1986; Miyamoto et al. 1991), since the launch of
\rxte~ the source has been monitored and complete sets of transitions
have been observed and studied (Belloni et al. 2005; Zdziarski et al.
2004; Del Santo et al. 2008). Unlike the BHC prototype Cyg X--1, for
which spectral state transitions are directly correlated with
luminosity, \gx\ shows hysteresis in its relation between X-ray
luminosity and spectral state \cite{zdz-gie04}, as also observed in
other BHC in low-mass binary systems (LMXB; Maccarone \& Coppi 2003;
Smith, Heindl \& Swank 2002). It was observed that hard-to-soft state
transitions during the rise phase occurs at higher luminosities than the
soft-to-hard transitions during the decline phase (Smith, Heindl \&
Swank 2002; Zdziarski et al. 2004).

{\it CGRO}, {\it Ginga} and \rxte\ data from \gx\ collected in the
period 1987-2004 have been analysed by Zdziarski et al. (2004). These
authors reported on long term variability and spectral correlations for
the $\sim$15 outbursts of \gx\ that occurred in this period.
Furthermore, in that work a lower limit for the source distance at 7 kpc
was provided.

The 2002/2003 outburst (Smith et al. 2002a; Nespoli et al. 2003; Buxton
\& Bailyn 2004) was followed with \rxte\ in detail through timing and
hardness-ratio (HR) analysis \cite{bellonial05}. These authors described
hysteresis in term of the source's evolution through a
hardness-intensity diagram (HID). Belloni et al. (2005) also introduced
new definitions for the different sub-states observed during state
transitions: the hard-intermediate (HIMS) and the soft-intermediate
states (SIMS; see also Homan \& Belloni 2005; Belloni 2005). For a
description of historical and alternative states classification
see Tanaka \& Lewin (1995), van der Klis (1995) and a more recent review
by McClintock \& Remillard (2006).

During the 2002/2003 outburst, close to the transition to the SIMS
(on 2002 May 17th \cite{smith02b}), a rapid (hours) bright radio flare
was observed \cite{fender02} which led to the formation of a large-scale
relativistic jet \cite{gallo04}. Fender, Belloni \& Gallo (2004)
associated this flare and subsequent matter ejections with the crossing
of the so-called jet-line, i.e.\ the HIMS-SIMS transition as reported by
Nespoli et al. (2003). During this HIMS-SIMS transition \gx\ showed fast
changes in its timing properties, but almost none in the 3-20 keV
energy spectrum  \cite{homan05}.

In February 2004 a new outburst started, which reached a significantly
lower peak flux than the 2002/2003 outburst (Buxton et al. 2004; Smith
et al. 2004; Belloni et al. 2004; Kuulkers et al. 2004). To get
broad-band coverage during the expected HIMS-SIMS spectral transition,
simultaneous \rxte\ and \integral\ observations were made. Belloni et
al. (2006) combined data from PCA, HEXTE and IBIS, and obtained good
quality broad-band (3-200 keV) energy spectra before and soon-after the
transition. These spectra indicated steepening of the hard, high-energy
component. Also, the  high-energy cut-off that was present at $\sim$70
keV before the transition was not detected later. Therefore, although
spectral parameters at lower energies do not change abruptly through the
transition, the energy of the cut-off increases or disappears rather
fast (within 10 hours). The power spectra before and after the
transition showed significant differences (see Belloni et al. 2005;
Belloni 2008): from strong band-limited noise and type-C QPO to much
weaker noise and type-B QPO (for a description of the properties of
different types of QPO, see Casella, Belloni \& Stella 2005).

In 2006 \rxte\ monitoring of \gx\ revealed low-level X-ray activity
(Bezayiff \& Smith 2006; Swank et al. 2006) until December, when a new
strong outburst started \cite{krimm06}. Miller et al. (2007) triggered a
public \integral\ ToO campaign on \gx\ which started on January
30th (Caballero-Garcia et al. 2007a-e). We activated our \rxte\
campaign with the aim to follow the source through a HIMS-SIMS
transition. Unfortunately, we did not observe the main transition, as we
did in 2004, but managed to capture a secondary transition. In
this paper, we report the results of the timing and spectral analysis of
our \rxte\ data from 2007 March 4--6 and of the quasi-simultaneous
interval of the \integral\ public data.

\begin{table*} 
\renewcommand{\arraystretch}{1.3}
\begin{tabular}{lccccccc} 
\hline 
\textbf{Interval} &\multicolumn{4}{c}{\bf{\integral}} &\multicolumn{3}{c}{\bf{\rxte}}\\
\hline 
& \textbf{Start (UT)} & \textbf{End (UT)} & \textbf{SCWs
interval}$^{\dag}$ &\textbf{Exp.}& \textbf{Start (UT)} & \textbf{End
(UT)} &\textbf{Exp.}\\

$\alpha$ & March 4 (17:51)  & March 4 (23:06) & 10-16 & 17.9 ks &  March
4 (16:45:36) &  March 4 (17:57:04) & 3.7 ks\\ 
$\beta$ & March 4 (23:08) & March 5 (21:09) & 17-45 & 73.5 ks &  March 5 (13:22:40) &  March 5
(14:18:40) & 3.1 ks \\ 
$\gamma$ & March 5 (21:35)  & March 6 (13:04) &
46-65 & 52.4 ks & March 6 (12:40:16)  &  March 6 (13:51:44) & 3.9 ks \\

\hline 
\end{tabular} 
\caption[]{\footnotesize Observing log of the three
time intervals used for the IBIS/ISGRI averaged spectra, as well as the
simultaneous \rxte~ pointings.\\ $\dag$ \integral\ Science Windows are
tagged by a number indicating the related orbit as well as the pointing
internal to it. Here we show the interval of the used pointings.}
\label{tab:log} 
\end{table*}

\begin{figure}
\begin{center}
\includegraphics[angle=90,width=8.5cm]{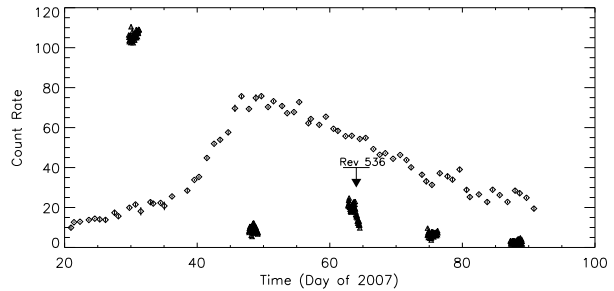}
\end{center} 
\caption{IBIS/ISGRI (20-40 keV) light curve of \gx\ (triangles) during
the \integral\ ToO campaign (Miller et al.\ 2007). Each point
corresponds to a single science window ($\sim$2500 s). The period
analysed in this work is marked. The \rxte/ASM count rate in the sum band
(2-12 keV) daily average is overplotted (diamonds).} 
\label{fig:lc_tot} 
\end{figure}

\section{Observations and data analysis} In order to follow the new
outburst of \gx\ at high energies, an \integral~ ToO campaign was
carried out. Starting from 2007 January 30 (MJD 54130), six observations
of 130 ks each (spaced by two weeks) were planned \cite{miller07}. A
total number of 255 pointings (5 observations), referred to as ``science windows'' (SCW),
lasting each roughly 2500 seconds, were performed. These observations
covered a two month period starting from \integral\ revolution (rev) 525
until rev 544. Our analysis focused on ISGRI \cite{lebrun03}, the low
energy detector layer of the coded mask imager, IBIS \cite{ube03}. Fig.
\ref{fig:lc_tot} shows the ISGRI light curve of the full \integral\
campaign. We report here on a short time interval, 56 SCWs of the orbit
536 (indicated in Fig. \ref{fig:lc_tot}), during which the transition
occurred (Tab. \ref{tab:log}).

\begin{figure} 
\begin{center}
\includegraphics[width=9cm]{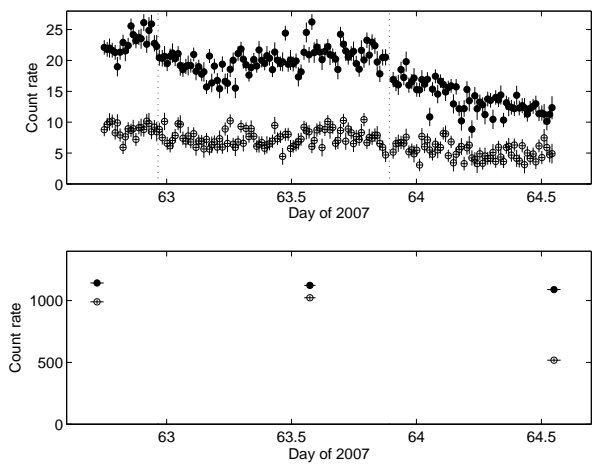} 
\end{center}
\caption{\integral\ and \rxte\ light curves of \gx\ during the period
2007 March 4.7--6.6. Top panel: IBIS/ISGRI light curves in the 20--40
keV ({\it filled circles}) and 40--80 keV ({\it open}) bands. The time
bin size is 1000 seconds. Dotted vertical lines separate the three IBIS
intervals used for the spectral analysis, i.e. $\alpha$ ({\it left}),
$\beta$ ({\it middle}) and $\gamma$ ({\it right}) (see text and Tab.
\ref{tab:log}). Bottom panel: \rxte\ light curves from PCA (2.5--20.2
keV, {\it filled circles}) and HEXTE (20--40 keV, {\it open circles}).
The HEXTE rates have been multiplied by 65 in order to allow for easier
comparison with the PCA rates. Each point corresponds to a single
observation. } 
\label{fig:lc} 
\end{figure}

The IBIS/ISGRI scientific analysis was performed using the \integral\
off-line analysis software, OSA 7.0. The total ISGRI light-curve with a
bin size equal to the duration of the SCW was obtained by extracting
counts rate of \gx\ from images (Fig. \ref{fig:lc_tot}). In Figure \ref{fig:lc} ({\it top})  
we show the ISGRI light-curves of rev 536 in the energy
ranges 20-40 keV and 40-80 keV. These count rates in a bin size of 1000 seconds
have been obtained with the OSA tool \texttt{ii\_lc\_extract}. Spectra were extracted from
each pointing with the \texttt{ii\_spectra\_extract} script in 35
logarithmic energy bins spanning from 20 keV to 1 MeV. These IBIS/ISGRI
spectra were averaged in three groups (Tab. \ref{tab:log}) and 1.4\% of
systematic errors were added.

In 2007, the \rxte\ campaign on \gx\ consisted of a large number of
observations covering the full outburst (Motta et al. 2008). For this
work, we isolated three pointings overlapping with the \integral\
observation periods (see Tab. \ref{tab:log} and Fig. \ref{fig:lc}, {\it bottom}).
We extracted PCA and HEXTE energy spectra (background and deadtime corrected) from each of
the three observations using the standard \rxte\ software within {\tt
HEASOFT} V. 6.4, following the standard extraction procedures. For
spectral analysis, only PCU2 from the PCA and  Cluster B from HEXTE have
been used. A systematic error of 0.6\% has been added to the PCA spectra
to account for residual uncertainties in the instrument calibration. We
accumulated background-corrected PCU2 rates in the
channel\footnote{corresponding to the original 0--128 channels}
bands $A$ = 6--48 (2.5--20.2 keV), $B$ = 6--14 (2.5--6.1 keV) and $C$ =
23--44 (9.4--18.5 keV). $A$ is the total rate, while the hardness was
defined as $H = C/B$ (see Homan \& Belloni 2005). For the timing
analysis of the PCA data we produced power spectra from 16-s stretches
accumulated in the channel band 0--35\footnote{corresponding to the
original 0--255 channels} (2--15 keV) with a time resolution of
1/128 s. This resulted in spectrograms of 243, 201 and 253 power spectra
for the three observations, respectively (see Nespoli et al. 2003). The
power spectra were normalised according to Leahy et al. (1983) and
converted to squared fractional rms (Belloni \& Hasinger 1990). For
different time selections (see below), we averaged the power spectra and
subtracted the contribution due to Poissonian noise (see Zhang et
al. 1995). The timing analysis was performed with custom software.

\begin{figure} 
\begin{center} \includegraphics[width=8cm]{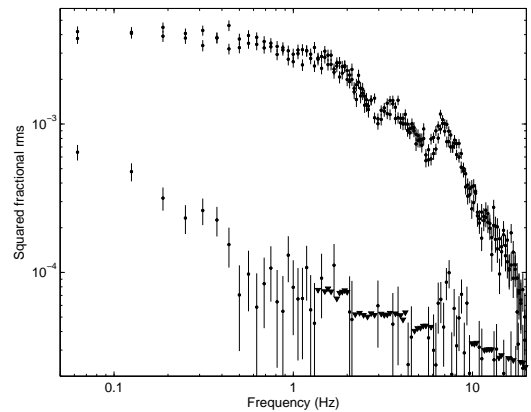}
\end{center} 
\caption{Power density spectra from PCA observations
$\alpha$, $\beta$ (top, overlapping) and $\gamma$ (bottom). The
presence of band-limited noise and a type-C QPO at $\sim$7 Hz is evident
from the top spectrum The bottom spectrum shows a lower level of noise
and a very weak type-A QPO around 8 Hz.} 
\label{fig:pds} 
\end{figure}

\section{Results}

\subsection{Evolution and timing analysis}

\gx\ light curves collected with the high-energy instruments, IBIS/ISGRI
and HEXTE (Fig. \ref{fig:lc}), show a count rate decrease by about a
factor of two from March 4.7 to 6.6 (days 62.7 to 64.6 of 2007).
Note that the IBIS count rate in the low/hard state at the beginning of
the outburst (30th January) was much higher (see Fig. \ref{fig:lc_tot}). Based on time
averaged count rates, we divided the IBIS data in three groups, referred
to as $\alpha$, $\beta$ and $\gamma$ (Tab. \ref{tab:log}, Fig.
\ref{fig:lc}). In the energy range 20--40 keV, we measured variations between groups $\alpha$ and
$\beta$, and between $\beta$ and $\gamma$, of about 25\% and 32\%,
respectively. Variations within each group are
less than 15\%. At low energies (2.5--20.2 keV), however, the
three PCA pointings show almost constant count rates, suggesting
that a transition occurred between groups $\beta$ and $\gamma$.

Timing analysis confirms this suggestion. The power density spectra
(PDS) corresponding to the PCA data of the three \rxte\ observations
indicate that observations $\alpha$ and $\beta$ are very similar, while
observation $\gamma$ is different (see Fig. \ref{fig:pds}). The PDS of the former
features  band-limited noise and a quasi-periodic oscillation (QPO)
around 7 Hz, i. e. type-C QPOs, with a Q factor (defined as the ratio
of the centroid frequency of the QPO by its FWHM) around 2.5.
The total integrated 0.1--64 Hz fractional rms is
$\sim$12\% (in 2--15 keV). The PDS of observation $\gamma$ corresponds
to much weaker variability ($\sim$2.4\%) with a hint of a broad excess
around 8 Hz, possibly a type-A QPO (Q$\sim$3). These results indicate that
observations $\alpha$ and $\beta$ correspond to a HIMS, while
observation $\gamma$ corresponds to a SIMS.

\begin{figure} \begin{center}
\includegraphics[angle=0,width=9cm, height=9cm]{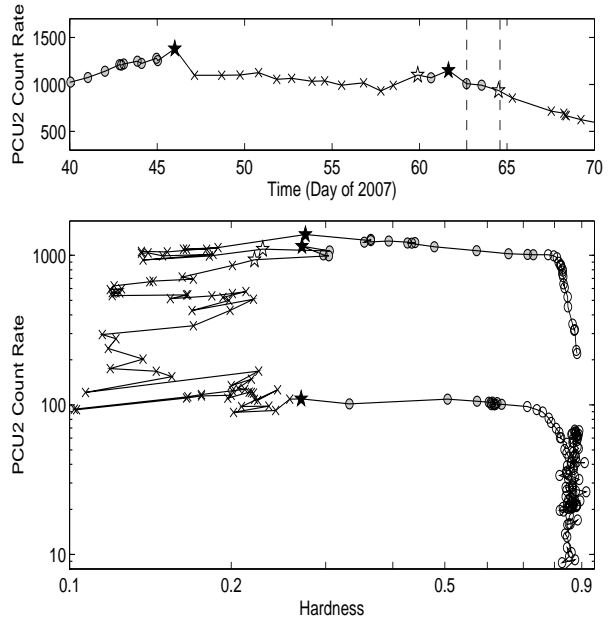} \end{center}
\caption{{\it Top}: PCU2 (2.5--20.2 keV) light curve of \gx\ since February
9 until March 11. These points (each of them represents a \rxte\
observation) correspond to the higher horizontal branch of the HID shown
below. Our observations lies between dashed lines; the symbols follow
the same convention of the figure below. {\it Bottom}:
Hardness-Intensity diagram for the complete 2006/2007 outburst which
starts from the middle right, around 200 cts/s, and proceeds in an
anticlockwise direction. Different symbols indicate timing properties
related with the presence of QPOs: type-A (white star), type-B (black
star), type-C (white circles), soft-state observations without little
variability and no QPO peaks (crosses). } \label{fig:hid} \end{figure}

In addition to our observations, the outburst has been followed in
detail with a public \rxte\  program, leading to a good coverage of the
full outburst. In order to describe the global behaviour of \gx\
in 2007, we show in Figure \ref{fig:hid} ({\it bottom}) the HID of all
available data, together with the PCU2 light curve (Fig. \ref{fig:hid},
{\it top}) spanning the one-month period encompassing our \integral\
observations. The symbols in Fig. \ref{fig:hid} indicate different
shapes of the power density spectra (see below; a full spectral/timing
analysis will be presented in a forthcoming paper).

Comparing this diagram with the HIDs previously reported for the
2002/2003 and 2004 outbursts (Belloni et al. 2005; Belloni et al. 2006),
we note a similar {\it q}-shaped track pattern as described in Homan \&
Belloni (2005; see also Belloni 2008). Between day 45 and day 46, a HIMS
to SIMS transition occurred (Motta et al. 2008). Moreover, at the
beginning of March \gx\ displayed additional HIMS/SIMS transitions, as
indicated by switches between various types of low-frequency QPOs (see
Figures \ref{fig:pds} and \ref{fig:hid}). From the symbols in Fig.
\ref{fig:hid} we can see that the observed sequence C-C-B
corresponds to the transition that occurred earlier (days 45/46), while
the transition reported here corresponds to the QPO sequence C-C-A (days 62.7/64.6).
These last three \rxte\ observations have been performed
simultaneously with \integral\ (see Fig. \ref{fig:lc}).

\begin{figure} 
\begin{center} \includegraphics[width=8cm,angle=90]{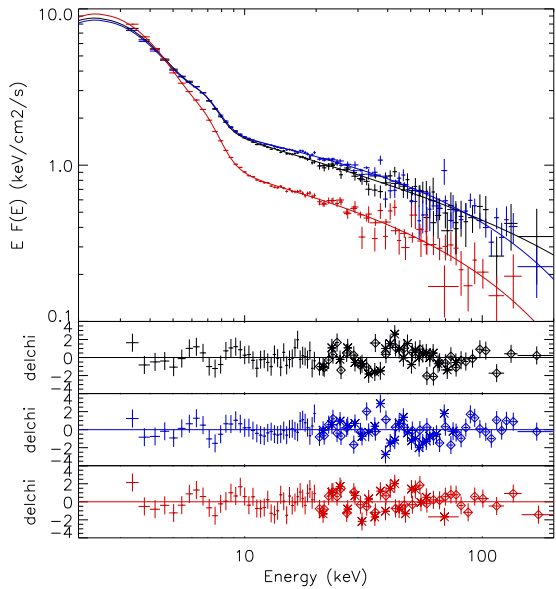} 
\end{center} 
\caption{Unfolded energy spectra with PCA, ISGRI and HEXTE, total models and residuals of groups $\alpha$
(black), $\beta$ (blue) and $\gamma$ (red) are shown. ISGRI and HEXTE residuals
are marked with diamonds and asterisks, respectively.} 
\label{fig:spec}
\end{figure}

\subsection{Spectral analysis}

We performed a broad-band spectral analysis of data collected with three
different instruments: PCA (3--20 keV), HEXTE (20--90 keV) and
IBIS/ISGRI (20--200 keV). Spectra were combined according to the
intervals defined in Section 3.1 (see Tab. \ref{tab:log} and Fig.
\ref{fig:lc}). {\tt XSPEC} version 11.2.3 was used for spectral fits.

A simple model consisting of multi-color disc-blackbody ({\sc{diskbb}})
plus a cut-off power law ({\sc{cutoffpl}}) was used to fit spectra shown in
Fig. \ref{fig:spec}. The hydrogen column density measured with
instruments having a low-energy coverage, e.g. \chandra, was taken
into account by adding a {\sc{wabs}} component to the model and freezing
$N_{H}$=5$\times 10^{21}$ cm$^{-2}$ (M\'endez \& van der Klis 1997; Kong
et al. 2000). An Iron emission line with centroid fixed at 6.4 keV was
needed to obtain good fits. To account for cross-calibration problems,
multiplicative constants of 0.9 and 1.1 for the HEXTE and IBIS/ISGRI
spectra (as compared to the PCA) were added to the fits. The above model
resulted in good fits (see Tab.\ref{tab:parameter}). Leaving the high
energy cut-off out of the model did not yield acceptable $\chi ^{2}$
values.

\begin{figure} 
\begin{center}
\includegraphics[width=9cm]{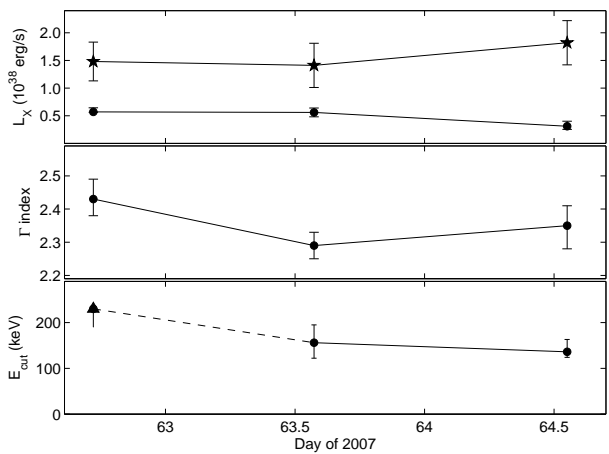} 
\end{center} 
\caption{Top panel: bolometric luminosities of the multicolor disc blackbody from
Stefan-Bolzmann law (filled stars) and 3-300 keV luminosities of the
power law with high-energy cut-off. The distance was assumed to be 8
kpc. Middle panel: best-fit power-law photon index. Bottom panel:
best-fit high-energy cut-off. The dashed line connects the lower limit
value found for interval $\alpha$ to the value of interval $\beta$. }
\label{fig:par} 
\end{figure}

At low energies we found the disc temperature (kT$_{bb}$) and inner
radius to remain constant within the errors (90\% confidence). However,
variations in the high-energy component included a decrease of the
power-law index (Fig. \ref{fig:par}, {\it middle}) and the high-energy
cut-off (Fig. \ref{fig:par}, {\it bottom}) between period $\alpha$ and
$\beta$.

Taking the best-fit values of radius and inner disc temperature
(Tab. \ref{tab:parameter}), we estimate the bolometric luminosity of the
disc component as 1.5, 1.4 and 1.8 $\times 10^{38}$ \lerg\ for the three
time intervals, respectively (Fig. \ref{fig:par}, {\it top}; stars).
While the disk flux increases by a factor of only 1.3, the decrease
in the high-energy component luminosity is about a factor of 1.8 (from
5.7 to 3.1$\times 10^{37}$ \lerg) from period $\beta$ to $\gamma$  (Fig.
\ref{fig:par}, {\it top}; filled circles).  The latter seems to be
mostly the result of the decrease in the power-law normalisation,
instead of the variation of the cut-off.

Based on observations performed in 2004 \cite{delsanto08}, we investigated a possible 
second non-thermal component, partially responsible for the high-energy spectra in HIM state.
Due to the short exposure time and statistics (spectra up to 200 keV), we did not manage to constrain parameters of an 
additional power-law on a thermal Comptonization component, as in Del Santo et al. (2008).

A further scenario is suggested by the presence of the iron emission line. We tried to model the  
high-energy component of our spectra with an exponentially cut-off power-law 
reflected from neutral material ({\sc pexrav}, Magdziarz \& Zdziarski 1995).
While the disc temperature is consistent with the previous  model, the high
energy component is significantly different (Tab. \ref{tab:pexrav}).
With the addition of a reflection component, we cannot constrain the
high energy cut-off, neither derive a lower limit. Moreover, the
reflection component becomes significant in the intervals 
$\beta$ and $\gamma$.  

\section{Discussion} 
Comparing the 2007 outburst of \gx\ with 
previous outbursts (from 2002/2003 and 2004) in terms of their HID tracks 
(see Fig. 2 in Belloni et al. 2006), we see similar {\it q}-shaped
tracks \cite{hm05}. The count rate level of the upper horizontal branch
in the 2007 is slightly higher than that of 2002/2003, and well above
the one of 2004 (more than a factor of 3). The evolution of states
throughout the outburst, as estimated from the timing properties, is
also very similar to the previous ones. The HIMS-SIMS transition on the
upper branch takes place at a similar color as the secondary transition
described in this work.

On 2007 March 4-6, we have observed a ``mini-transition'' from the HIMS to SIMS
in \gx, which occurred on a secondary horizontal branch of the 
{\it q}-track (Fig. \ref{fig:hid}). During this secondary transition, clearly
marked by changes in the properties of fast aperiodic timing (from
band-limited noise and type-C QPO to power-law noise and a weak type-A
QPO), an important change in the power-law flux component was
observed. This seems to be consistent with results from Nespoli et al.
(2003) and Belloni et al. (2005) suggesting that power-law flux and
timing are tightly correlated during the fast transitions.

Moreover, a change in the high energy cut-off and photon index appeared between the two
HIMS observations rather than between the HIMS and the SIMS. 
During this secondary transition we observed a slight delay between variations of the spectral parameters and 
the ones of the hard-X flux and timing properties.

The power-law with high energy cut-off ($E_{c}$) is a simple model to
describe thermal Comptonization of soft photons by a hot electron
plasma. A decrease in $E_{c}$ may indicate the  
cooling of the electrons plasma, possibly causing the energy reservoir for
thermal Comptonization to go down. 
The increase of disc luminosity causing the cooling of the hot corona 
is usually observed in hard-to-soft state transitions of BHC
(Zdziarski \& Gierli\'nski 2004; Del Santo et al. 2005; Del Santo et al. 2008). 
During the SIM state described in this work, we found indications for a slight increase (a factor of $\sim$1.3) in the disc 
flux, simultaneously with a decrease of the high energy flux by a factor of two.

An alternative scenario for explain the delay between cut-off and flux changes,
involves the presence of a second, possibly non-thermal, source of high-energy photons.
As the corona cools (between intervals $\alpha$ and $\beta$), the second high energy component
continues to be responsible for most of the flux. After the transition HIMS-to-SIMS this second component 
disappears, resulting in a decrease of the high energy flux.

In 2004, the major HIMS to SIMS transition on the primary horizontal
branch was simultaneously observed with \integral\ and \rxte\
\cite{belloni06}. After the transition, these authors report on the lack
of the high energy cut-off in the SIMS (present at $\sim$70 keV in the
HIMS). Del Santo et al. (2008) confirm the latter result (disappearance
of the cut-off in the SIMS) by using simultaneous IBIS, SPI and JEM-X
data collected during the same transition. However they found a higher
value of the cut-off in that same HIM state ($115_{-23}^{+27}$ keV)
because of new \integral\ calibrations. Here, we observe a secondary
HIMS-to-SIMS transition with a different behaviour: during the transition
the high-energy cut-off has moved to a lower energy. Moreover, this
variation is observed to take place {\it before} the transition as
deduced from the timing properties, although we cannot exclude additional
very fast timing transitions between the three intervals.

Also, a fit with a {\sc{pexrav}} model fails to measure a significant
cut-off for any of the intervals examined here. In this scenario the
spectral changes observed between intervals $\alpha$ and $\beta$ could
be due to the appearance of the Compton reflection component.
The increase of the reflection component, as the spectrum becomes softer, is 
expected in the truncated disc model \cite{done07}.

In summary, the transitions observed in 2004 and 2007 display different
properties at high energies. Other differences which might play a role
are:

\begin{itemize} 
\item the 2004 outburst peaked at a considerably lower luminosity (and count rate) than that of 2007; 
\item the transition in 2004 corresponded to the main transition from hard to soft, while in
2007 we observed a secondary transition; 
\item after the transition, in 2004 a type-B QPO was observed, while in 2007 a type-A QPO was observed.
\end{itemize}

These observations clearly indicate that the properties of the
high-energy component(s) in the spectrum of \gx\ during a HIMS-to-SIMS
transition are complex. Del Santo et al. (2008) analysed a uniform
\integral\ data-set collected during the 2004 outburst, including the
fast transition presented in Belloni et al. (2006). These authors report
on different state transitions: HIMS-to-SIMS-to-HSS.
All these transitions, occurring when \gx\ was going forward (from right to left) on the main
horizontal branch, have been explained as being driven by increase 
of the soft cooling photon flux in the corona. 
The transition HIMS-to-SIMS presented in this work (in 2007) 
occurred on the secondary-branch of the {\it q}-pattern and showed different behaviour,
even though it occurred at same color as the 2004 transition.

In conclusion, we still do not have a complete picture of the evolution of
the hard spectral component during the whole transition
LHS to HSS. More coordinated observations, such as those
presented here, are needed.

\section*{Acknowledgments} MDS is supported by the Italian
Space Agency (ASI), via contract INTEGRAL I/008/07/0. MDS thanks Julien Malzac
for useful scientific discussion.
As one of the thousands Italian researchers with a long-term temporary
position, MDS acknowledges the support of Nature (455, 835-836) and thanks
the Editors for increasing the international awareness of the current
critical situation of the Italian Research. 
MDS and AB aknowledges the \integral\ data archival support at IASF-Roma by Memmo Federici.
TMB acknowledges support from the International Space Science Institute
(ISSI) and from ASI via contract I/088/06/0.
We thank J. Miller et al. for immediately making the data from their \integral\
program publicly available.

\begin{table*} 
\renewcommand{\arraystretch}{1.5}
\begin{tabular}{lccccccc} 
\hline Interval &\multicolumn{3}{c}{Disc blackbody} &\multicolumn{3}{c}{Cut-Off
power-law}& $\chi ^{2}_{\nu}$ (d.o.f.)\\ 
\hline 
& \textbf{kT [keV]} & \textbf{R$_{in}$ [km]} & \textbf{F$_d$ [erg cm$^{-2}$s$^{-1}$]} &
\textbf{$\Gamma$} & \textbf{E$_c$ [keV]} & \textbf{F$_p$ [erg
cm$^{-2}$s$^{-1}$]} \\ 
\hline

$\alpha$ & 0.84$^{+0.01}_{-0.03}$   & 48$^{+4}_{-2}$ &
4.2$_{-0.3}^{+0.9} \times 10^{-9}$ &2.43$^{+0.06}_{-0.05}$ & $>$230 &
7.5$_{-0.9}^{+0.8}\times 10^{-9}$& 1.07(99)\\

$\beta$ & 0.83$^{+0.03}_{-0.02}$  & 48$\pm 4$ & 4.3$^{+0.7}_{-0.3}
\times 10^{-9}$ &2.29$\pm 0.04$ & 156$_{-34}^{+39}$ & 7.4$^{+0.3}_{-1.1}
\times 10^{-9}$ & 0.99 (99)\\

$\gamma$ & 0.85$^{+0.01}_{-0.02}$  & 52$^{+2}_{-1}$ & 5.5$_{-0.2}^{+0.1}
\times 10^{-9}$ &2.35$_{-0.07}^{+0.06}$ & 136$^{+27}_{-12}$ &
4.1$_{-0.7}^{+0.5} \times 10^{-9}$ & 1.06 (99)\\

\hline 
\end{tabular} 
\caption[]{\footnotesize Spectral fitting
parameters for the three time intervals obtained using an absorbed
($N_{H}$ frozen at $5 \times 10^{21}$ cm$^{-2}$) multicolor disc
black-body (with temperature as kT) plus a power-law ($\Gamma$) with
high energy cut-off ($E_{c}$). For R$_{in}$, the assumed distance and
inclination are 8 kpc and 50$^\circ$ \cite{zdz04}. The fluxes of the two
model components, namely black-body (F$_d$) and cut-off power-law
(F$_p$), have been calculated in the 3-20 keV and 3-300 keV bands,
respectively. Errors represent 90\% confidence limits. A 6.4 keV
Gaussian line was also required for the fits. }
\label{tab:parameter}
\end{table*}

\begin{table*} 
\renewcommand{\arraystretch}{1.5}
\begin{tabular}{lcccc} 
\hline 
Interval &  kT [keV] & $\Gamma$ & $\Omega$/${2\pi}$ & $\chi ^{2}_{\nu}$
(d.o.f.)\\ 
\hline

$\alpha$ & $0.84^{+0.02}_{-0.03}$   & $2.52^{+0.03}_{-0.14}$ & $< 0.2$ &
1.09(98)\\

$\beta$ & $0.82^{+0.03}_{-0.02}$   & $2.51^{+0.07}_{-0.12}$ &
0.4$^{+0.2}_{-0.1}$ & 1.10(98)\\

$\gamma$ & $0.84^{+0.02}_{-0.01}$  & $2.65^{+0.04}_{-0.19}$ &
$0.5^{+0.1}_{-0.2}  $ & 1.0(98)\\

\hline 
\end{tabular} 
\caption[]{\footnotesize Results of spectral fits
with {\sc{pexrav}} plus {\sc{diskbb}}. A Fe line a 6.4 keV was modeled
with {\sc gauss}.} 
\label{tab:pexrav} 
\end{table*}

\label{lastpage}

\end{document}